\documentclass[aps,prl,twocolumn,showpacs]{revtex4}

\usepackage[dvips]{graphicx}
\usepackage{amssymb}

\newcommand{\eps} {\varepsilon}

\begin{document}

\title{Coarse Graining in Block Copolymer Films}
\author{Yoav Tsori }
\affiliation{Department of Chemical Engineering, Ben-Gurion University of the
Negev, \\P.O. Box 653, 84105 Beer-Sheva, Israel}

\author{David Andelman}
\affiliation{School of Physics and Astronomy\\ Raymond and Beverly Sackler
Faculty of Exact Sciences\\ Tel Aviv University, Ramat Aviv, Tel
Aviv 69978 Israel}

% version submitted to cond-mat}
%\date{\today}

\begin{abstract}

We present few ordering mechanisms in block copolymer melts in the
coarse-graining approach. For chemically homogeneous or modulated
confining surfaces, the surface ordering is investigated above and
below the order-disorder temperature. In some cases the copolymer
deformation near the surface is similar to the copolymer morphology
in bulk grain boundaries. Block copolymers in contact with rough
surfaces are considered as well, and the transition from lamellae
parallel to perpendicular to the surface is investigated as a
function of surface roughness. Finally, we describe how external
electric fields can be used to align block copolymer meso-phases in
a desired direction, or to induce an order-order phase transition,
and dwell on the role of mobile dissociated ions on the transition.

\flushleft{{\bf  keywords: block copolymers, confinement, electric fields,
phase transitions}}
\end{abstract}

%\pacs{}

\maketitle

%\twocolumn

%\newpage

%
%
%\newpage
%\baselineskip=24pt
%%%%%%%%%%%%%%%%%%%%%%%%%%%%%%%%%%%%%%%
\section{Introduction}
%%%%%%%%%%%%%%%%%%%%%%%%%%%%%%%%%%%%%%%

Block copolymers (BCP) are heterogeneous polymers where each polymer
chain is composed of several chemically distinct homopolymer blocks,
connected together by a covalent bond. These polymeric systems exhibit
fascinating structures in the nanometer scale and can be created by
self-assembly from solutions or the melt state \cite{hamley1,fred}.

In addition, BCP are composite materials that have many applications. For
example, by  connecting together a stiff (rod-like) block with a flexible
(coil) block, one can obtain a material which is rigid, but not brittle
\cite{yachin1,yachin2}. Moreover, the interplay between flexibility and
toughness can be controlled by temperature. Different chain architecture
(ring or star-like) may lead to novel mechanical and flow properties
\cite{satkowski}. In addition, BCP have many industrial uses because the
length scales involved are smaller or comparable to the wavelength of
light. These applications include waveguides, photonic band gap materials
and other optoelectronic devices \cite{VBSN01} and dielectric mirrors
\cite{FT98}.

Our prime concern in this mini-review are melts of BCP above the
glass transition. However, it is worthwhile mentioning that BCP also
exhibit interesting properties upon cooling below the glass
transition into a solid state. Some BCP may undergo crystallization
of one or more components that is accompanied by strong structural
changes, while other BCP systems stay in the vitrified state upon
cooling.

In the molten state, because of competition between enthalpy and
entropy, at high temperatures the BCP melt behaves as a disordered
fluid, while at low temperatures the macroscopic phase separation is
hindered because the two (or more) immiscible sub-chains cannot be
detached from each other as they try to phase separate due to block
incompatibility. Hence, BCP'S phase separate into a variety of
micro-ordered structures, with characteristic size depending on the
BCP chain length and other system parameters \cite{hamley1,fred}.
The morphology and structure of the prevailing phase depend on the
lengths of constituent sub-chains (also called blocks), the chemical
interactions between the blocks, the temperature and the chain
architecture. The BCP micro-domain size ranges from about $10$ to
several hundreds nanometers.

A typical example of a well-studied di-BCP, polystyrene-polyisoprene
(PS-PI), is shown in Fig. 1 \cite{khandpur95}. As temperature is
cooled below the order-disorder temperature (ODT), the disorder melt
of chains micro-phase separates into one of the meso-phases:
lamellar, hexagonal, body centered cubic (bcc) or gyroid.

%fig1
\begin{figure}[h!]
\begin{center}
\includegraphics[scale=0.5,bb=60 290 535 810,clip]{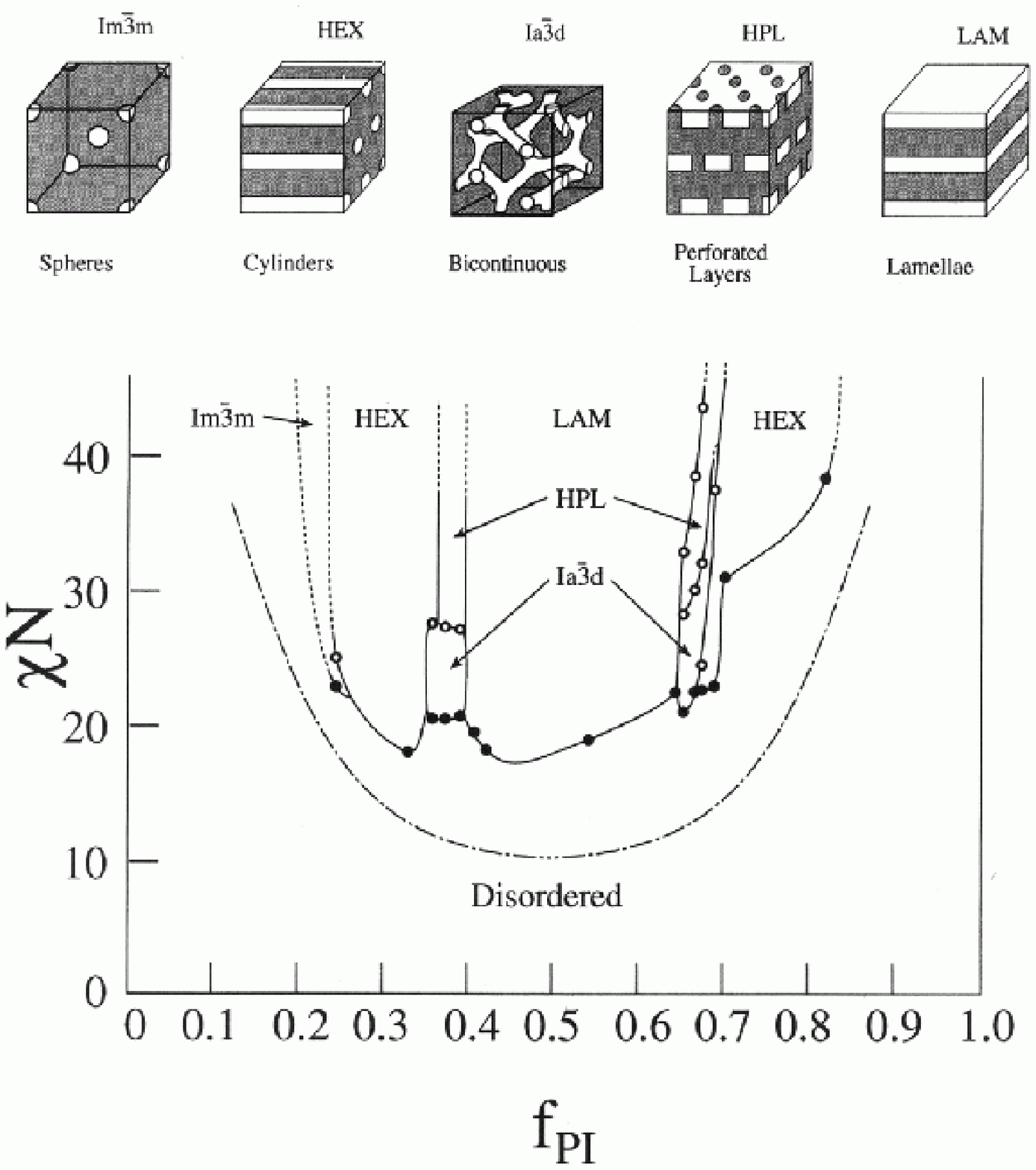} 
\end{center}
\caption{\textsf{ $\chi N$ versus $f_{\rm PI}$ phase diagram for
PI-PS diblock copolymers. The dash-dot curve is the mean field
prediction for the ODT. Solid curves have been drawn to delineate
the different phases observed but might not correspond to precise
phase boundaries. Five different ordered microstructures (shown
schematically) have been observed for this chemical system. Adapted
with permission from ref~\cite{khandpur95}.}}
\end{figure}

The present paper reviews  several mechanisms that can be used to
achieve a desired ordering  and orientation in thin films of BCP.
The simple analytical model is presented in Sec. \ref{model}, valid
in the weak segregation limit (WSL). We consider in Sec.
\ref{above_odt} BCP's above the ODT point and in contact with
chemically patterned surfaces. The polymer density is given as a
function of a pre-designed and fixed chemical pattern on the
surface.
Thin films of di-BCP's below the ODT confined between two flat and
parallel surfaces are investigated in Sec. \ref{below_odt}. We find
that for a one-dimensional chemical surface pattern, the lamellae
are tilted with respect to the parallel surfaces.  If the surface
and lamellar periodicities are equal, the lamellae are formed
perpendicular to the surfaces. We relate  the orientation phenomenon
to the formation of tilt boundary (chevrons) defects in bulk
lamellar phases.
Beside chemically heterogeneous surfaces, rough surfaces are investigated
in Sec. \ref{rough}, and a simple explanation to the parallel to
perpendicular transition in these system as function of surface roughness
is proposed. Alignment of confined lamellae by external electric fields
is studied in Sec. \ref{E_field}. It is shown that because different
polymers have different values of the dielectric constant,  the
electrostatic energy favors an orientation of lamellae in a direction
perpendicular to the confining electrodes. This electrostatic tendency
can be used to overcome interfacial interactions with the bounding
electrodes and align structures in a desired direction.

%%%%%%%%%%%%%%%%%%%%%%%%%%%%%%%%%%%%%%%%%%%%%
\section{The model}\label{model}
%%%%%%%%%%%%%%%%%%%%%%%%%%%%%%%%%%%%%%%%%%%%%

Let us consider first a simple free energy expression for  A/B
di-BCP melt.  With the definition of the order parameter $\phi({\bf
r})\equiv\phi_A({\bf r})-f$ as the local deviation of the A monomer
concentration from its average, the bulk free energy can be written
as \cite{Leibler80,binder1}:
\begin{eqnarray}\label{F}
\frac{Nb^3F_b}{k_BT}=\int\left\{\frac12\tau\phi^2+\nonumber\right. \\
\left.\frac12h\left(
\nabla^2\phi+q_0^2\phi\right)^2
+\frac16\Lambda\phi^3+\frac{u}{24}\phi^4\right\}{\rm d}^3r
\end{eqnarray}
$d_0=2\pi/q_0$ is the fundamental periodicity in the system, and is
expressed by the polymer radius of gyration $R_g$ through $q_0\simeq
1.95/R_g$. The parameter $\tau=2\rho N\left(\chi_c-\chi\right)$
measures the distance from the critical point ($\tau=0$) in terms of
the  Flory parameter $\chi\sim 1/T$. At the critical point (or
equivalently the ODT) $\chi_c\simeq 10.49/N$. In addition, $b$ is
the Kuhn statistical segment length, $h=1.5\rho c^2R_g^2/q_0^2$ and
$\rho=1/Nb^3$ is the chain density per unit volume. $\Lambda$ and
$u$ are the three- and four-point vertex functions calculated by
Leibler \cite{Leibler80}.

For simplicity we will restrict most (but not all) the discussion
below to lamellar phases of symmetric BCP'S. This allows us to
simplify the above free energy by considering only the symmetric
case:  $f=\frac12$ and  the cubic $\Lambda$-dependent term drops
out. In addition, we set for convenience $c=u/\rho=1$ throughout the
paper. However, in Sec. \ref{E_field} where we treat the bcc to
hexagonal transition we will consider also the cubic term in the
free energy as it is indispensable to describe asymmetric phases.

BCP's \cite{FH87,OK86,TAepl01} and other systems with spatially
modulated phases \cite{Swift77} have been successfully described by
Eq.~(\ref{F}) or similar forms of free energy functionals. The free
energy, Eq.~(\ref{F}), describes  a system in the disordered phase
having a uniform $\phi=0$ for $\chi<\chi_c$ (positive $\tau$), while
for $\chi>\chi_c$ (negative $\tau$), the system is in the lamellar
phase for $f=\frac12$, $\Lambda=0$, and can be described
approximately by a single $q$-mode $\phi=\phi_L\exp(i{\bf q_0}\cdot
{\bf r})$. The amplitude of the sinusoidal modulations is given by
$\phi_L^2=-8\tau/u$. The validity of Eq.~(\ref{F}) is limited to a
region of the phase diagram close enough to the critical point where
the expansion in powers of $\phi$ and its derivatives is valid, but
not too close to it, because then critical fluctuations become
important \cite{B-F90,braz1}. This limit employed hereafter is
called the weak segregation limit (WSL).

%%%%%%%%%%%%%%%%%%%%%%%%%%%%%%%%%%%%%%%%%%%%%%%
\section{Disordered BCP's in contact with chemically patterned
surfaces}\label{above_odt}
%%%%%%%%%%%%%%%%%%%%%%%%%%%%%%%%%%%%%%%%%%%%%%%

The presence of chemically heterogeneous but otherwise flat surface
is modeled by adding short-range surface interactions to the free
energy of the form,
\begin{equation}\label{Fs}
F_s=\int\left[\sigma({\bf r_s})\phi({\bf r_s}) +\tau_s\phi^2({\bf
r_s})\right]{\rm d^2}{\bf r_s}~~~+~~~const.
\end{equation}
The integration is carried out over the position of the confining
surfaces parameterized by the vector ${\bf r_s}$. The surface field
is $\sigma({\bf r_s})\equiv \gamma_{_{\rm AS}}-\gamma_{_{\rm BS}}$,
where $\gamma_{_{\rm AS}}$ and $\gamma_{_{\rm BS}}$ are the
interfacial interactions of the A and B blocks with the surface,
respectively. Furthermore, $\sigma$ has an arbitrary but fixed
spatial variation and is coupled linearly to the BCP surface
concentration $\phi({\bf r_s})$. Preferential adsorption of the A
block ($\phi>0$) onto the {\it entire} surface is modeled by a
constant $\sigma<0$ surface field, resulting in parallel-oriented
layers (a perpendicular orientation of the chains). One way of
producing such a surface field in experiments is to coat the
substrate with random copolymers \cite{L-RPRL96,M-RPRL97}. If the
pattern is spatially modulated, $\sigma({\bf r_s})\neq 0$, then the
A and B blocks are attracted to different regions of the surface.
The coefficient of the $\phi^2$ term in Eq.~(\ref{Fs}) is taken to
be a constant surface correction to the Flory parameter $\chi$
\cite{F87,TAjcp01}. A positive $\tau_s$ coefficient corresponds to a
suppression of surface segregation of the A and B monomers.

%%%%%%%%%%%%%%%%%%%%%%%%%
\subsection{One surface}
%%%%%%%%%%%%%%%%%%%%%%%%

 For simplicity we consider first BCP'S confined by one surface located at
$y=0$ as is depicted in Fig. 2 (a). A generalization to two parallel
surfaces is not difficult and will be given later. The surface chemical
pattern $\sigma({\bf r}_s)=\sigma(x,z)$ can be decomposed in terms of its
$q$-modes
%
%fig 2
\begin{figure}[h!]
\begin{center}
\includegraphics[scale=0.45,bb=25 470 570 640,clip]{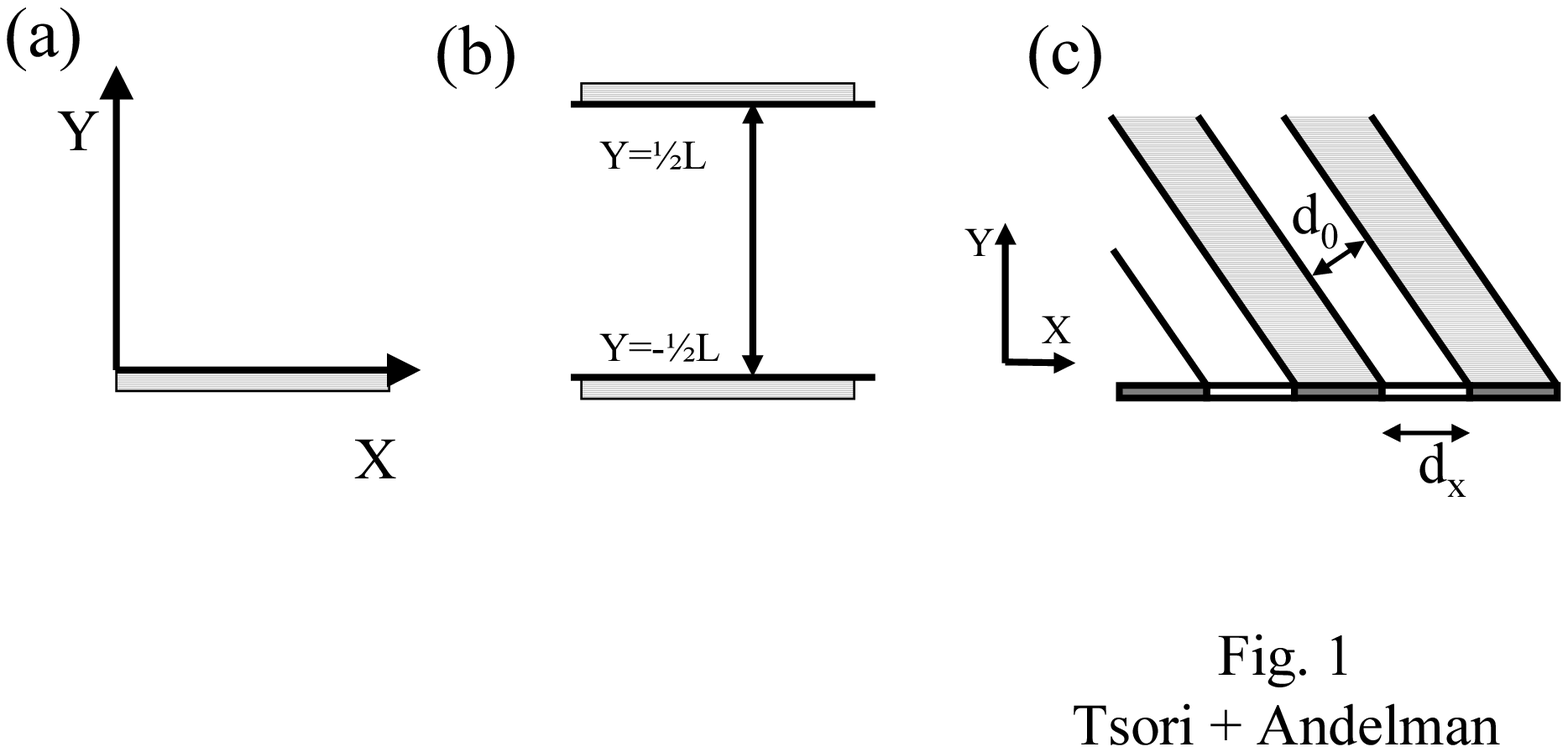} %25 405 570 640
\end{center}
\caption{\textsf{ Schematic illustration of the coordinate system
for BCP'S confined by one [part (a)] or two [part (b)] planar and
parallel surfaces. (c) Lamellae are formed tilted with respect to
the surface if the surface periodicity $d_x$ is larger than the
natural one $d_0$.}}
\end{figure}

\begin{equation}
\sigma(x,z)= \sum_{\bf q} \sigma_{\bf q}{\rm e}^{i(q_xx + q_zz)}
\end{equation}
where ${\bf q}=(q_x,q_z)$, and $\sigma_{\bf q}$ is the mode
amplitude. Similarly, $\phi$ can be written as a sum
\begin{equation}
\phi({\bf r})=\sum_{\bf q}\phi_{\bf q}(y){\rm e}^{i(q_xx+q_zz)}
\end{equation}
Close to the ODT the free energy is stable to second order in
$\phi$, and higher order terms (i.e. the $\phi^4$ term) can be
neglected. Then $\phi({\bf r})$ is inserted into Eq. (\ref{F}) and
an integration over the $x$ and $z$ coordinates is carried out.
Minimization with respect to $\phi_{\bf q}(y)$ yields the
Euler-Lagrange equation
\begin{equation}\label{EL-fq}
\left[\tau/h+\left(q^2-q_0^2\right)^2\right]\phi_{\bf q}
+2(q_0^2-q^2)\phi_{\bf q}^{\prime\prime}+\phi_{\bf
q}^{\prime\prime\prime\prime}=0
\end{equation}
Note that the equation is linear and that the Fourier harmonics
$\phi_{\bf q}$ are not coupled. The boundary conditions are rather
complicated because they couple the value of the amplitude and its
derivatives at the surface. They result from minimization of the
full free energy expression, Eqs.~(\ref{F}) and (\ref{Fs})
\begin{eqnarray}\label{bc1}
\phi_{\bf q}^{\prime\prime}(0)+(q_0^2-q^2) \phi_{\bf q}(0)&=&0~~~~~\\
 \sigma_{\bf q}/h+2\tau_s\phi_{\bf q}(0)/h+(q_0^2-q^2)\phi_{\bf
q}^{\prime}(0)+\phi_{\bf q}^{\prime\prime\prime}(0) &=&0~~~~~\label{bc2}
\end{eqnarray}
Since Eq. (\ref{EL-fq}) is linear, its solution is a sum of
exponentials,
\begin{eqnarray}
\phi_{\bf q}(y)=A_{\bf q}\exp(-k_{\bf q}y)+B_{\bf q}\exp(-k^*_{\bf
q}y)
\end{eqnarray}
where the modulation constant $k_{\bf q}$ and the amplitude $A_{\bf
q}$ are given by
\begin{eqnarray}
k_{\bf q}^2 &=&q^2-q_0^2+i\sqrt{\tau/h}\label{kq}\\ \nonumber A_{\bf
q}&=&-\sigma_q\left(4\tau_s+2{\rm Im}(k_{\bf q})\sqrt{\tau
h}\right)^{-1}
\end{eqnarray}
In the above ${\rm Re}(k_{\bf q})>0$ ensuring that $\phi_{\bf q}\to
0$ as $y\rightarrow\infty$. This restricts the solution $\phi_{\bf
q}$ to be a sum of only two (out of four) exponential terms.

The two lengths, $\xi_q=1/{\rm Re}(k_{\bf q})$ and $\lambda_q=1/{\rm
Im}(k_{\bf q})$, correspond to the exponential decay and oscillation
lengths of the ${q}$-modes, respectively. For fixed $\chi$, $\xi_q$
decreases and $\lambda_q$ increases with increasing $q$. Close to the ODT
and for  $q>q_0$  we find finite $\xi_q$ and
$\lambda_q\sim(\chi_c-\chi)^{-1/2}$. However, all $q$-modes in the band
$0<q<q_0$ are equally ``active'', i.e., these modes decay to zero very
slowly in the vicinity of the ODT as $y\rightarrow\infty$:
$\xi_q\sim(\chi_c-\chi)^{-1/2}$ and $\lambda_q$ is finite. Therefore, the
propagation of the surface imprint (pattern) of $q$-modes with $q<q_0$
into the bulk can persist to long distances, in contrast to surface
patterns with $q>q_0$ which persist only close to the surface. The
$q=q_0$ mode has both lengths $\xi_q, \lambda_q$ diverging as
$(\chi_c-\chi)^{-1/4}$ for $\chi\rightarrow\chi_c$.

%fig 3
\begin{figure}[h!]
\begin{center}
\includegraphics[scale=0.45,bb=15 300 580 475,clip]{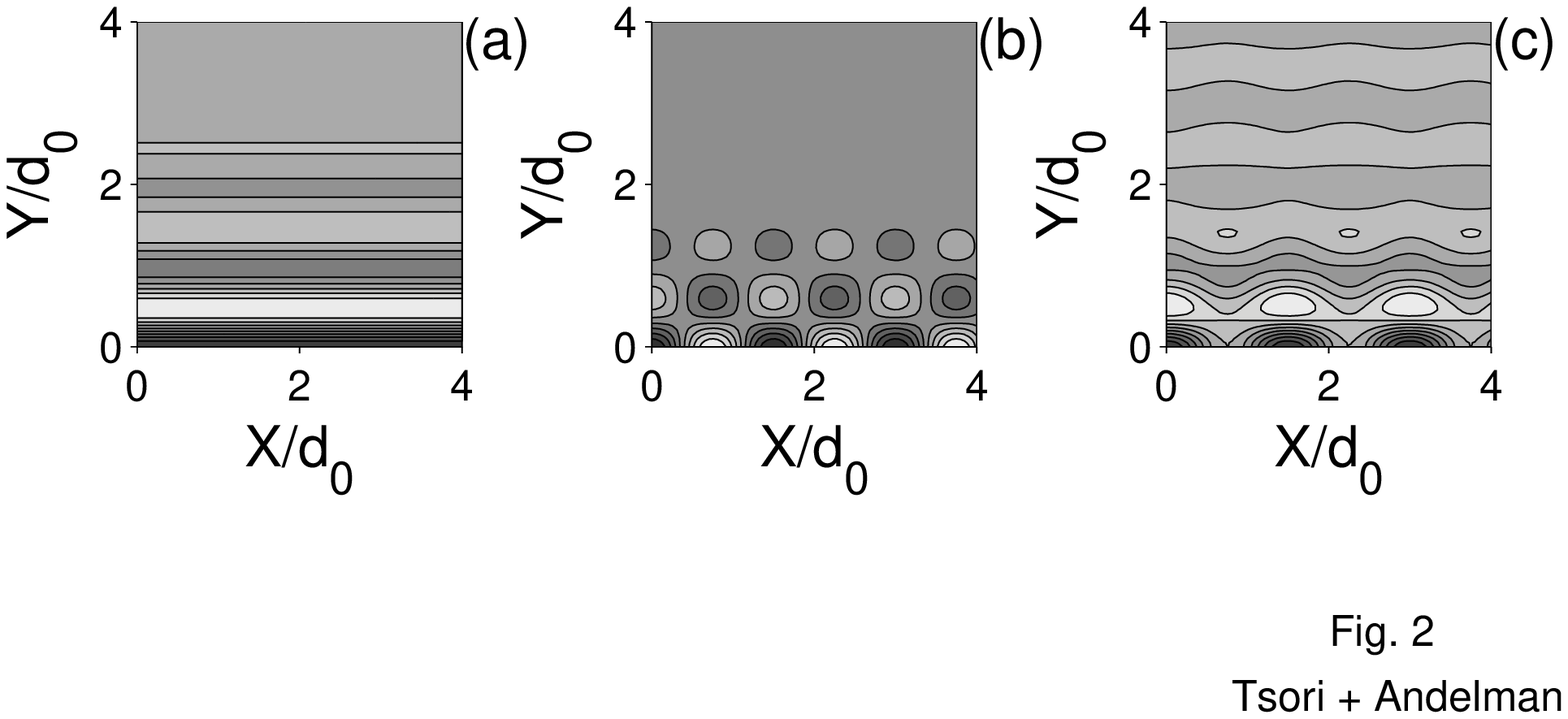} %15 240 580 475
\end{center}
\caption{\textsf{A BCP melt confined by one surface at $y=0$.
B-monomer density is high in dark regions, while A monomers are in
light regions. In (a) the surface is uniform, $\sigma=0.3$ and in
(b) it has stripes given by $\sigma=0.3\cos(\frac23q_0)$. The
``combined'' effect is shown in part (c) where
$\sigma=0.3+0.3\cos(\frac23q_0x)$ has a uniform and modulated part.
The Flory parameter is $N\chi=10.2$, $\tau_s=0$ and lengths in the
$x$ and $z$ directions are scaled by the lamellar period $d_0$.
Adapted from ref~\cite{TAIntSci03}. }}
\end{figure}

In Fig.~3 we give examples of the polymer morphologies in the case of
three simple surface patterns. A uniform surface [in (a),
$\sigma=\sigma_0$ is constant] causes exponentially decaying density
modulations to propagate in the $y$-direction. A striped surface [in (b),
$\sigma=\sigma_q\cos(qx)$] creates a disturbance that is periodic in the
$x$-direction, which decays exponentially in the $y$-direction. The
combined surface pattern [in (c), $\sigma=\sigma_0+\sigma_q\cos(qx)$]
induces density modulations which are the sum of the ones in (a) and (b).

A more complex chemical pattern, shown in Fig.~4 (a), consists of V
shaped stripes on the $y=0$ surface. The polymer density in parallel
planes with increasing distance from the surface is shown in (b) and (c).
Note how the frustration induced by the tips of surface chemical pattern
[in (a)] is relieved as the distance from the surface increases.
Similar morphology is observed when two grains of lamellar
phase meets with a tilt angle, creating a tilt grain boundary in bulk
systems.

%fig 4
\begin{figure}[h!]
\begin{center}
\includegraphics[scale=0.48,bb=15 300 550 490,clip]{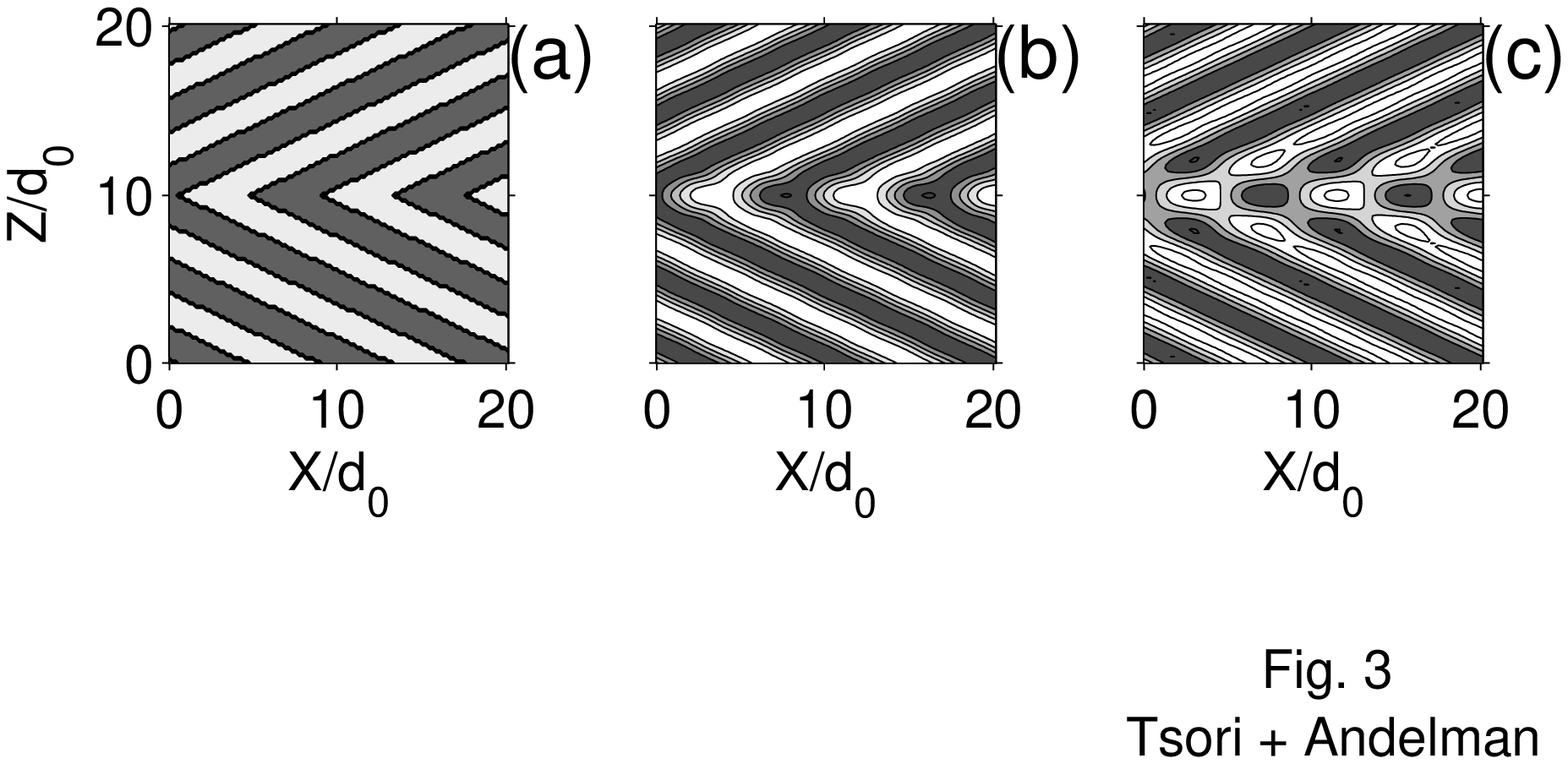} %15 248 550 490
\end{center}
\caption{\textsf{ Propagation of surface pattern into the bulk. The
surface pattern in the $y=0$ plane is shown in (a), where white
(black) show regions preferring A (B) monomers. Parts (b) and (c)
are contour plots of the polymer density at $y=3d_0$ and $y=8d_0$,
respectively. $N\chi=9.5$ and $\tau_s=0$. Adapted from
ref~\cite{TAIntSci03}. }}
\end{figure}

%%%%%%%%%%%%%%%%%%%%%%%%%%%%
\subsection{Two confining surfaces}
%%%%%%%%%%%%%%%%%%%%%%%%%%%%

Our treatment of confined BCP'S can  be easily generalized to the
case of two flat parallel surfaces \cite{TAmm01}. The governing
equation is still Eq. (\ref{EL-fq}), but now there are four boundary
conditions instead of the two in Eqs. (\ref{bc1}) and (\ref{bc2}).
Figure~5 shows how two simple surface patterns can be used to
achieve a complex three-dimensional polymer morphology, even though
the melt is in its bulk disordered phase. The stripes on the two
surfaces are rotated by 90$^\circ$  with respect to each other. A
symmetric ``checkerboard'' morphology appears in the mid-plane.

%fig 5
\begin{figure}[h!]
\begin{center}
\includegraphics[scale=0.5,bb=45 270 570 490,clip]{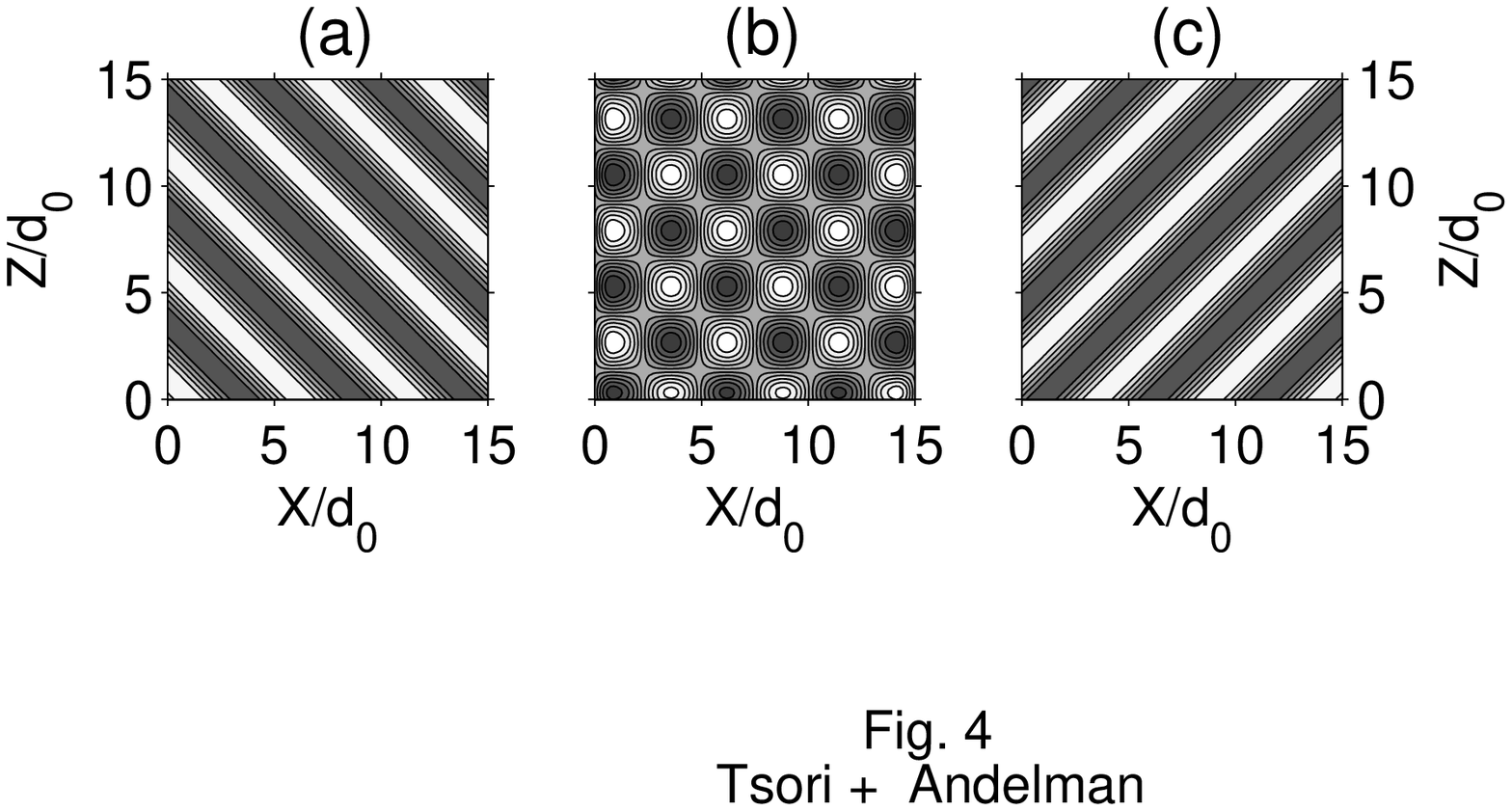} %45 229 570 490
\end{center}
\caption{\textsf{BCP melt confined by two flat parallel striped
surfaces, depicted in parts (a) and (c), and located at $y=-d_0$ and
$y=d_0$, respectively. The melt morphology in the mid-plane ($y=0$)
is shown in part (c). The Flory parameter is $N\chi=9$ and
$\tau_s=0$. Adapted from ref~\cite{TAmm01}.  }}
\end{figure}

Up to this point, the BCP melt was assumed to be in its bulk disordered
phase (above the bulk ODT point). When a melt in the lamellar phase
(below ODT) is confined in a thin film, the morphology is dictated by a
complex interplay between the natural periodicity and the imposed film
thickness.

%%%%%%%%%%%%%%%%%%%%%%%%%%%%%%%%%%%%%%%%%%%%
\section{Lamellar BCP between two chemically pattered surfaces}\label{below_odt}
%%%%%%%%%%%%%%%%%%%%%%%%%%%%%%%%%%%%%%%%%%%%

In this section we describe the ordering of lamellar BCP's confined
by one or two surfaces. The phase behavior of thin BCP films in the
lamellar phase subject to {\it uniform} surface fields has been
investigated experimentally \cite{W-RMM94} and theoretically
\cite{F87,turnerPRL92,matsen1,matsen2,pbmm97,G-M-B00,wang1}, and was
found to consist of parallel, perpendicular and mixed lamellar
phase. The latter phase has parallel lamellae extending from one
surface, which are jointed in a T-junction defect with perpendicular
lamellae extending from the opposite surface \cite{P-WMM99,wang2}.
At a given inter-surface spacing, increasing the (uniform) surface
interactions promotes a parallel orientation with either A-type or
B-type monomers adsorbed onto the surface. However, if the spacing
$L$ between the surfaces is incommensurate with the lamellar
periodicity, or the incompatibility $\chi$ is increased, a
perpendicular orientation is favored \cite{TAepje01}.

In the treatment given below, a new effect can be observed when the
surfaces are taken to be non-uniform, ``striped'', with regions of
alternating preferences to the A and B blocks [see Fig.~2 (c)]
\cite{krausch0}. The stripe periodicity $d_x$ is assumed to be larger
than the natural (bulk) periodicity, $d_x>d_0$, and the stripes are
modeled by
\begin{equation}\label{sigma2}
\sigma(x,z)=\sigma_q\cos(q_xx)
\end{equation}
and are translational invariant in the $z$-direction. The surface
$q$-mode is $q_x=2\pi/d_x<q_0$.

Contrary to the system above the ODT, a linear response theory
assuming small order parameter as a response to the surface field is
inadequate here, since the bulk phase has an inherent spatially
varying structure. The surface effects are contained in the
correction to the order parameter
\begin{equation}
\delta\phi({\bf r}) \equiv\phi({\bf r})-\phi_b({\bf r})
\end{equation}
where $\phi_b$ is a ``tilted'' bulk lamellar phase given by
\begin{eqnarray}  \label{bulk}
\phi_b&=&-\phi_L\cos\left(q_xx+q_yy\right)\\
q_x&=&q_0\cos\theta,\qquad q_y=q_0\sin\theta,
\end{eqnarray}
The bulk ordering is depicted schematically as tilted lamellae in
Fig.~2 (c). For the correction order parameter $\delta\phi$ we
choose
\begin{equation}\label{dphi}
\delta\phi(x,y)=g(y)\cos(q_xx).
\end{equation}
This correction describes a lamellar ordering perpendicular to the
surface, and commensurate with its periodicity $d_x=2\pi/q_x$. The
overall morphology of the lamellae is a superposition of the
correction field $\delta\phi$ with the tilted bulk phase, having a
periodicity $d_0$. The region where the commensurate correction
field $\delta\phi$ is important is dictated by the amplitude
function $g(y)$. The total free energy $F=F_b+F_s$ is now expanded
about its bulk value $F[\phi_b]$ to second order in $\delta \phi$.
The variational principle with respect to $g(y)$ yields a master
equation:
\begin{equation}
\label{gov_geqn} \left[ A+C\cos(2q_yy)
\right]g(y)+Bg^{\prime\prime}(y) +g^{\prime\prime\prime\prime}(y)=0,
\end{equation}
with parameters $A$, $B$ and $C$ given by:
\begin{eqnarray}
A=-\tau/h+q_y^4~,~~~~~B=2q_y^2~,~~~~~ C=-\tau/h~.
\end{eqnarray}
This linear equation for $g(y)$ is similar in form to Eq. (\ref{EL-fq})
describing the density modulation of a BCP melt in the disordered phase.
The lamellar phase is non-uniform and this results in a $y$-dependency of
the term in square brackets. The above equation is readily solved using
the proper boundary conditions (for more details see refs.
\cite{TAjcp01,TASpre00}).

In Fig.~6 we present results for a BCP melt confined by one
sinusoidally patterned surface, $\sigma(x)=\sigma_q\cos(q_xx)$, with
no average preference to one of the blocks,
$\langle\sigma\rangle=0$, for several values of surface periodicity
$d_x$ and for fixed value of the Flory parameter $\chi>\chi_c$. The
main effect of increasing the surface periodicity $d_x$ with respect
to $d_0$ is to stabilize tilted lamellae, with increasing tilt
angle. Note that even for $d_x=d_0$ [Fig.~5a] yielding no tilt, the
perpendicular lamellae have a different structure close to the
surface as is induced by the surface pattern. Although the surface
interactions are assumed to be strictly local, the connectivity of
the chains causes surface-bound distortions  to propagate into the
bulk of the BCP melt. In particular, this is a strong effect in the
weak-segregation regime we are considering.

%fig 6
\begin{figure}[h!]
\begin{center}
\includegraphics[scale=0.5,bb=20 310 560 485,clip]{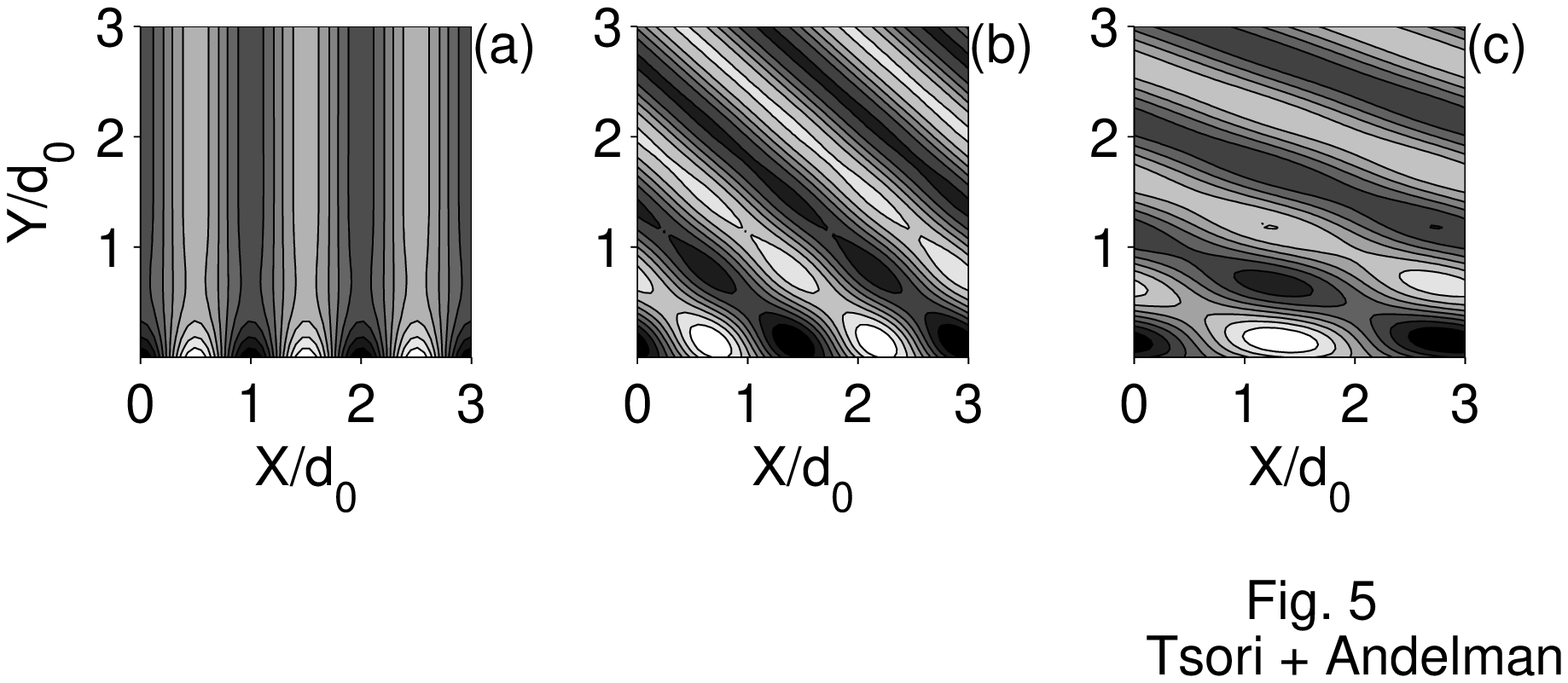} %20 273 560 485
\end{center}
\caption{\textsf{Tilted lamellar phase in contact with one patterned
surface at $y=0$. (See also Fig.~2c). The surface patterning is
modeled by the term $\sigma_q\cos(2\pi x/d_x)$. The lamellae tilt
angle $\theta=\arccos(d_0/d_x)$ increases as the periodicity of the
surface $d_x$ increases: $\theta=0$ for $d_x=d_0$ in (a),
$\theta\simeq 48.1^o$ for $d_x=\frac32 d_0$ in (b) and
$\theta\approx 70.5^o$ for $d_x=3 d_0$ in (c). In the plots
$\sigma_q/hq_0^3\phi_L=1$. The Flory parameter $N\chi=11.5$ and
$\tau_s/hq_0^3=0.1$. Adapted from ref~\cite{TAjcp01}. }}
\end{figure}

So far in this section we have considered the semi-infinite problem
of a BCP melt confined by one patterned surface. It is of
experimental and theoretical interest to study thin films of BCP'S
when they are confined between a heterogeneous (patterned) surface
and a second chemically homogeneous surface. This situation is
encountered when a thin BCP film is spread on a patterned surface.
The second interface is the film/air interface and is homogeneous.
Usually the free surface has a lower surface tension with one of the
two blocks. This bias can be modeled by adding a constant $\sigma_0$
term to the $\sigma(x)$ surface field. For simplicity, we assume
that the surface at $y=-\frac12 L$ has purely sinusoidal stripes
while at $y=\frac12 L$ the surface is attractive to one of the A/B
blocks with a constant preference:
\begin{eqnarray}
\sigma(x)&=&\sigma_q\cos(q_xx), \qquad\mbox{at \qquad $y=-\frac12
L$}
,\nonumber\\
\sigma(x)&=&\sigma_0, \qquad\qquad\qquad\mbox{at \qquad $y=\frac12
L$.}
\end{eqnarray}
A neutral surface at $y=\frac12 L$ is obtained as a special case
with $\sigma_0=0$.
%The striped surface pattern is at $y=-\frac12 L$ and
The expression (\ref{bulk}) for the bulk tilted phase is modified
($y \rightarrow y+\frac12 L$) in order to match the stripe surface
pattern at $y=-\frac12 L$,
\begin{equation}\label{bulk2}
\phi_b=-\phi_L\cos\left[q_xx+q_y(y+\frac12 L)\right]
\end{equation}
The homogeneous surface field at $y=\frac12L$ induces a lamellar
layering parallel to the surface, since the two A/B blocks are
covalently linked together. The simplest way to account for this
layering effect is
 to include an $x$-independent term $w(y)$ in
our ansatz, Eq.~(\ref{dphi}), for the order parameter:
\begin{equation}\label{dphi2}
\delta\phi(x,y)=g(y)\cos(q_xx)+w(y).
\end{equation}

The tilted lamellar phase confined by one homogeneous and one patterned
surface is a generalization of the mixed (perpendicular and parallel)
lamellar phase., which
occurs when the surface imposed periodicity $d_x$ is equal to the bulk
periodicity $d_0$. This ``T-junction'' morphology, shown in Fig.~7, has
perpendicular lamellae extending from the patterned surface. The
homogeneous field at the opposite surface favors a parallel orientation
of the lamellae. The crossover region between the two orientations is
found in the middle of the film, and its morphology depends on
temperature (the $\chi$ parameter). The effect of the homogeneous field
is evident, as parallel ordering extends from the top surface. We see
here that strong enough modulated surface fields stabilize the tilted
lamellar phases and, in particular, the mixed phase.

%fig 7
\begin{figure}[h!]
\begin{center}
\includegraphics[scale=0.6,bb=57 265 470 680,clip]{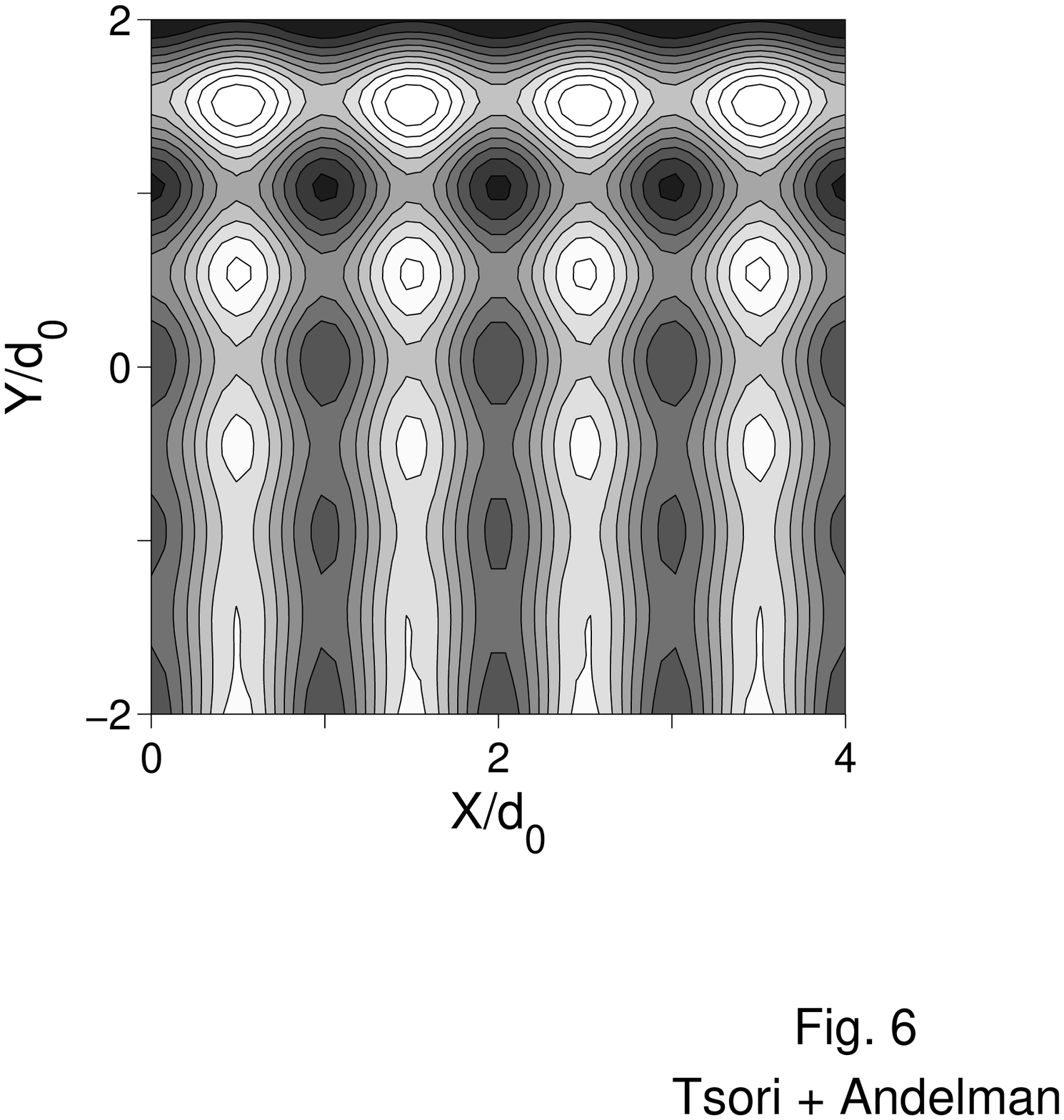} %50 163 580 680
\end{center}
\caption{\textsf{ A BCP confined film showing a crossover from
perpendicular lamellae at the $y=-\frac12 L=-2d_0$ surface to
parallel lamellae at the other surface, $y=\frac12 L$. The pattern
on the bottom surface, $\sigma(x)=\sigma_q\cos(q_0x)$, has the bulk
periodicity $d_0$, and amplitude $\sigma_q=2hq_0^3$, while the top
surface ($y=\frac12 L$) is homogeneously attractive to the B polymer
(in black), $\sigma_0=4hq_0^3$. The Flory parameter is given by
$N\chi=10.7$ and $\tau_s/hq_0^3=0.4$. Adapted from
ref~\cite{TAjcp01}.}}
\end{figure}

%Until this point, the BCP's have been assumed to be confined by
%ideally flat surfaces. We now turn to describe lamellar phases in
%the presence of rough surfaces.

%%%%%%%%%%%%%%%%%%%%%%%%%%%%%%%%%%%%%%%%%%
\section{BCP's in contact with rough surfaces}\label{rough}
%%%%%%%%%%%%%%%%%%%%%%%%%%%%%%%%%%%%%%%%%%

In the preceding sections we have described how the copolymer
morphology is influenced by chemically homogeneous or patterned
surfaces that are smooth and flat. Another way to control BCP
structure is by the use of rough or corrugated surfaces. This method
has some advantages because it is rather straightforward to
construct experimentally such surfaces \cite{rough1}. When a
lamellar stacking is placed parallel to a rough, sinusoidally
modulated, surface (see Fig. 8), the lamellar state is a compromise
between interfacial interactions preferring that the lamellae
closely follow the surface contour, and bending and compression
energies, preferring flat layers.

The surface roughness is modeled by a single one-dimensional corrugation
mode, whose height in the $z$-direction above an $(x,y)$ reference plane
is given by $h(x)=R\cos(q_sx)$. $q_s$ and $R$ are the wavenumber and
amplitude of the surface roughness, respectively (see Fig.~8). The BCP is
put above the substrate in the half-space $z\geq h(x)$. We denote
$\gamma_{_{\rm AB}}$ as the interfacial interaction between the A and B
blocks in the polymer chain.
%
%fig 8
\begin{figure}[h!]
\begin{center}
\includegraphics[scale=0.5,bb=75 370 505 580,clip]{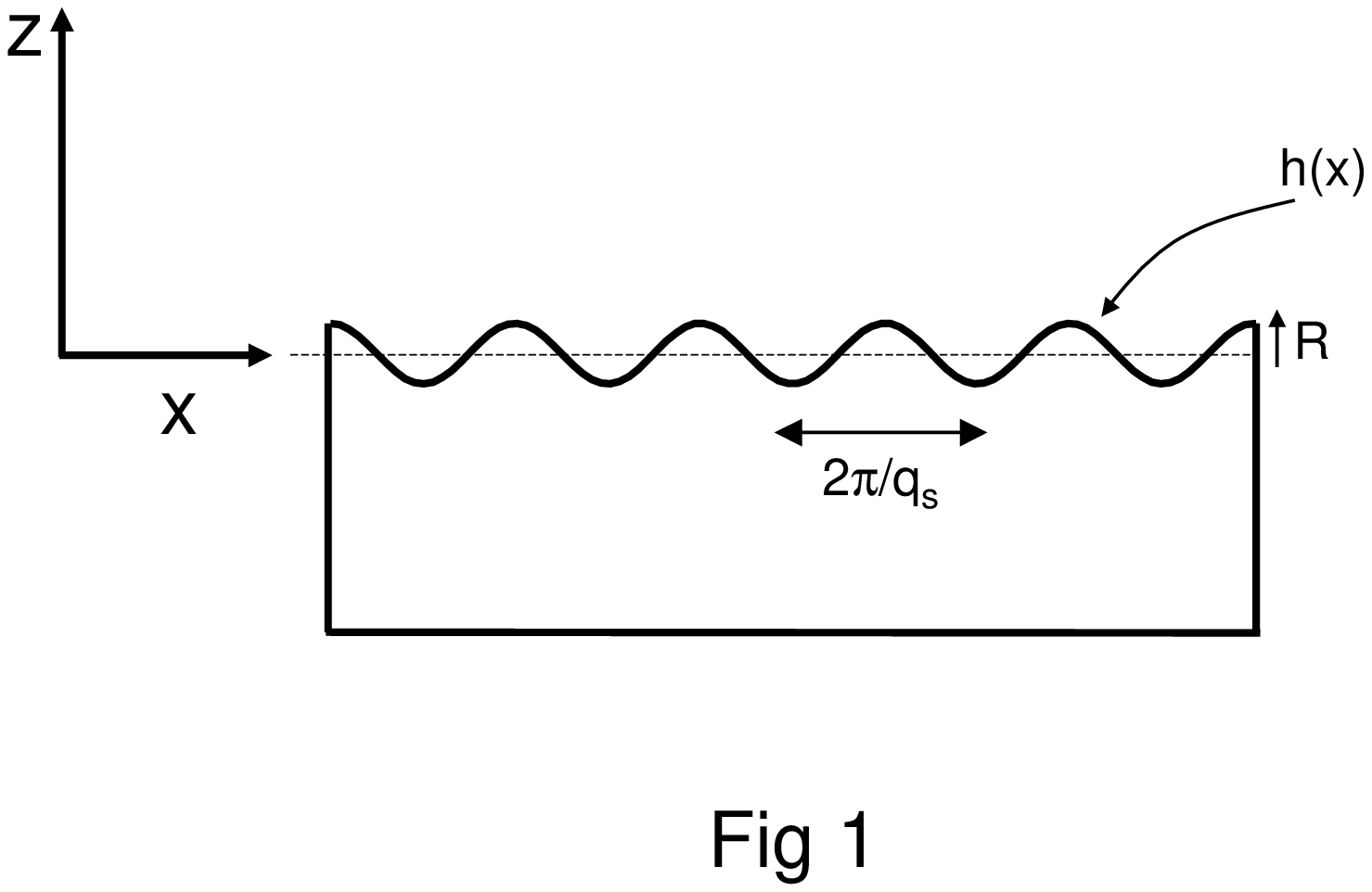}
\end{center}
\caption{\textsf{Schematic illustration of a rough surface, with
sinusoidal height undulations $h(x)=h_0+R\cos(q_sx)$.}}
\end{figure}

For lamellae oriented perpendicular to the surface, the order-parameter is
written as \cite{rough1}:
\begin{eqnarray}\label{phi_perp}
\phi_\perp({\bf r})=\phi_{0}\cos\left(q_0x+q_0u(x,z)\right)
\end{eqnarray}
The function $u(x,z)$ describes the surface-induced deviation of
the A/B interface from its flat (perfect) shape. The bulk part of
the free energy can be written as \cite{rough2,TJ,PGGP}:
\begin{eqnarray}\label{Fpara_u}
F_b=\frac12\int\left[K\left(u_{zz}\right)^2+B\left(u_x\right)^2\right]
{\rm d}^3r
\end{eqnarray}
where $u_x=\partial u/\partial x$, $u_{zz}=\partial^2u/\partial
z^2$, $K\sim d_0\gamma_{_{\rm AB}}$ is the bending modulus and
$B\sim\gamma_{_{\rm AB}}/d_0$ is the compression modulus.
To the elastic energy integral above must be added a term taking into account
the interfacial energies of the A and B blocks. This is simply given by
$\frac12\left(\gamma_{_{\rm AS}}+\gamma_{_{\rm BS}}\right)$, multiplied by a
correction factor. This factor is $1+\frac14\left(q_sR\right)^2$, reflecting
the fact that the real surface area is larger than the projected one, and
assuming small roughness $q_sR\ll 1$.
Minimization of the free-energy above gives the expression for $u$.
The deformation $u$ is larger close to
the surface, as can be seen from the deformation of the perpendicular lamellae
in Figure 9 (a).

Substitution of the expression for $u$ back into the free-energy
integral Eq. (\ref{Fpara_u}) gives the free-energy per unit area of the
perpendicular lamellae
\begin{eqnarray}\label{Fperp}
F_\perp\simeq\phi_0^2\frac{\left(\gamma_{_{\rm AS}}-\gamma_{_{\rm
BS}}\right)^2}{q_0K}+
\frac12\left(\gamma_{_{\rm AS}}+\gamma_{_{\rm BS}}\right)
\left(1+\frac14\left(q_sR\right)^2\right)\nonumber\\
\end{eqnarray}

The deformation $u$  for parallel lamellae can be achieved in a
similar way. The resulting parallel layering is seen in Fig.~9 (b),
with the same parameters as in part (a). The total free-energy in
this case is:
\begin{eqnarray}\label{Fpara}
F_{_\parallel}&\simeq&\phi_0^2\frac{\left(\gamma_{_{\rm AS}}-\gamma_{_{\rm
BS}}\right)^2}{q_0K}
\left(\frac{q_0}{q_s}\right)^2\left(q_0R\right)^2+
\left[\left(\frac12-\phi_0\right)\gamma_{_{\rm AS}}\right. \nonumber\\
&&\left. +\left(\frac12+\phi_0\right)\gamma_{_{\rm BS}} \right]
\left(1+\frac14(q_sR)^2\right)
\end{eqnarray}
Here again we find the same factor $1+\frac14\left(q_sR\right)^2$, but the
energies of interaction with the surface are different from the previous case:
the A and B polymers do not cover the surface equally, and hence
$\gamma_{_{\rm AS}}$ and $\gamma_{_{\rm BS}}$ have different prefactors in
the square brackets.
%
%fig 9
\begin{figure}[h!]
\begin{center}
\includegraphics[scale=0.8,bb=145 215 390 600,clip]{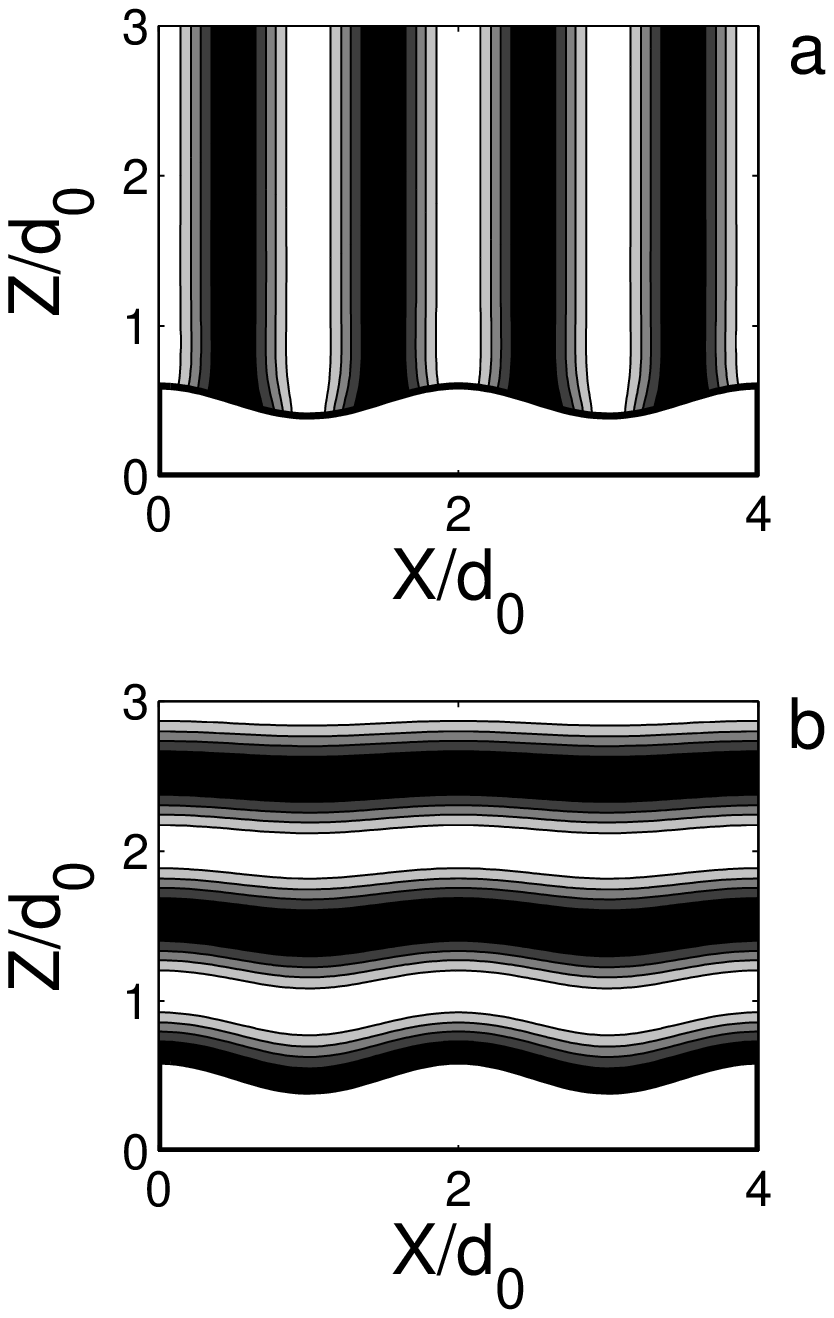}
\end{center}
\caption{\textsf{Perpendicular (a) and parallel lamellae (b) on
rough surfaces. The lamellar periodicity is half the surface one.
The parameter used are $B=2 \times 10^5$J/m$^2$, $K=B/(4q_0^2)$ and
$\sigma=\gamma_{_{\rm AS}}-\gamma_{_{\rm BS}}=\sqrt{BK}/4$. Adapted
from ref~\cite{rough1}.}}
\end{figure}

Based on the free-energies above, Eqs. (\ref{Fperp}) and
(\ref{Fpara}), a phase diagram can be constructed in the phase space
of three variables: the surface and lamellar inverse periodicities
$q_s$ and $q_0$, respectively, and the surface amplitude $R$.
%
%fig 10
\begin{figure}[h!]
\begin{center}
\includegraphics[scale=0.8,bb=190 65 420 725,clip]{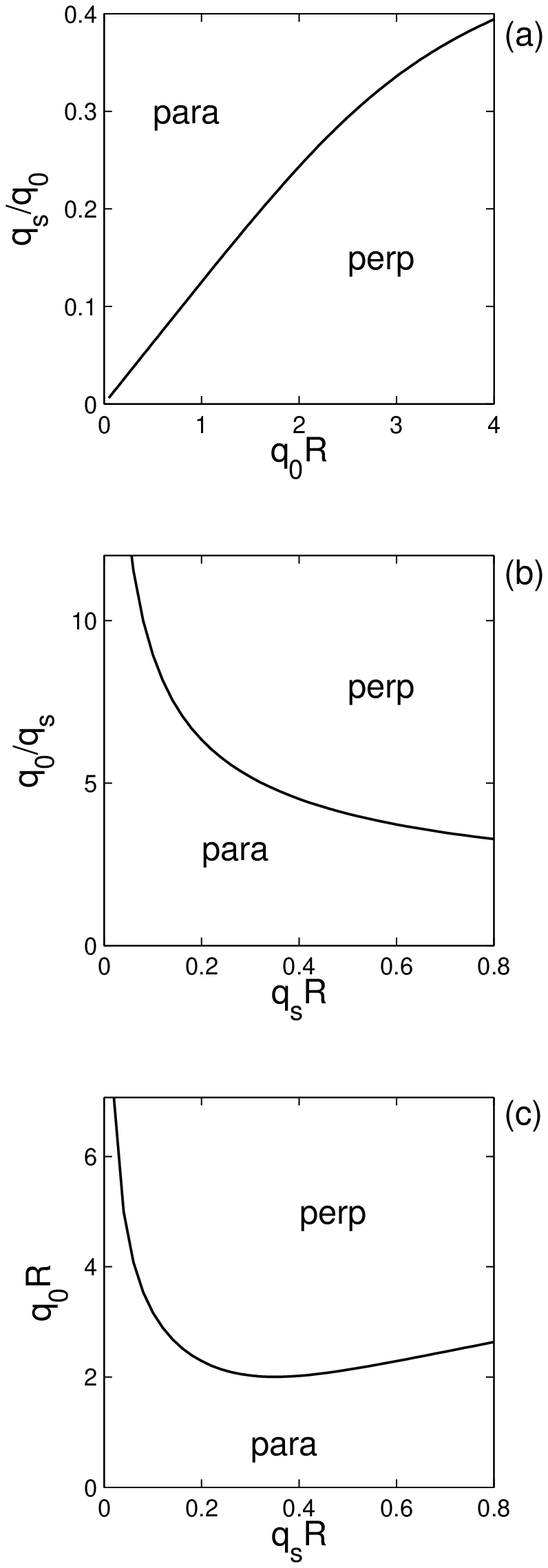}
\end{center}
\caption{\textsf{Phase-diagrams for perpendicular and parallel
lamellar on roughs substrates. (a) $R$ and $q_s$ are scaled by
$q_0$. (b) The surface wavenumber $q_s$ is used to scale $q_0$ and
$R$. (c) $q_s$ and $q_0$ are scaled by the surface amplitude $R$. In
the three plots we used $\phi_0=0.4$ The parameters used are
$\sigma=\gamma_{_{\rm AS}}-\gamma_{_{\rm BS}}=0.25$\,mN/m and
$\gamma_{_{\rm AB}}=1$\,mN/m. Adapted from ref~\cite{rough2}}}
\end{figure}
Three cuts in the phase diagram are given in Fig.~10. In (a), $R$
and $q_s$ are scaled by $q_0$. An increase in $q_s$ while keeping
$q_0$ and $R$ constant generally leads to a preference of parallel
layering. A different view is presented in (b), where $q_0$ and $R$
are scaled by $q_s$. Here, keeping $q_s$ and $q_0$ fixed while
increasing $R$ leads to a preference of perpendicular ordering.
Similarly, in (c) $R$ is used to scale $q_s$ and $q_0$. An increase
of $q_0$ at constant $q_s$ and $R$ favors perpendicular lamellae.

In the preceding sections we have considered ordering
mechanisms where
the interaction of the polymers with the confining surfaces is
mediated to regions far from the surfaces because of chain
connectivity. We now turn to discuss orientation of BCP films in
presence of external electric fields. This is a bulk ordering
mechanism that does not originate from the surface.

%%%%%%%%%%%%%%%%%%%%%%%%%%%%%%%%%%%%%%%%%%
\section{BCP's in presence of electric fields}\label{E_field}
%%%%%%%%%%%%%%%%%%%%%%%%%%%%%%%%%%%%%%%%%%

The influence that an electric field has on anisotropic polarizable
media (e.g. block copolymers) is of great importance. We will, in
particular,  concentrate on two aspects: orientational transitions
and order-to-order phase transitions.

%%%%%%%%%%%%%%%%%%%%%%%%%%%%%%%%%%%%%%%%%%%%
\subsection{Orientation of anisotropic phases by an electric field}
%%%%%%%%%%%%%%%%%%%%%%%%%%%%%%%%%%%%%%%%%%%%

When a material with inhomogeneous dielectric constant is placed in an
electric field $E$, there is an electrostatic free energy penalty for
having dielectric interfaces perpendicular to the field
\cite{TAmm02,AH93,AH94,Onuki95,PW99,AM01,TDR00,fraaije02}. This is the
so-called ``dielectric mechanism'' for BCP orientation. Thus, a state
where $\nabla\eps$ is perpendicular to the field ${\bf E}$ is favored
\cite{krausch1}.
The strength of this effect is proportional to $(\eps_A-\eps_B)^2E^2$,
where $\eps_A$ and $\eps_B$ are the dielectric constants of the polymers,
and is enhanced when the difference in polarizabilities is large.

Consider BCP in the lamellar phase and confined between two flat and parallel
electrodes. The lamellae may order parallel to the surfaces and suffer
some unfavorable stretching or compression, in order to gain better surface
coverage, as is discussed in the preceding sections. An applied electric field
perpendicular to the surface will tend to orient the lamellae parallel to
it, provided that it can overcome the interfacial interactions.
In the weak-segregation regime ($\phi\ll 1$), the
electrostatic energy per unit volume is given by the Amundson-Helfand
approximation \cite{AH93,AH94}
\begin{eqnarray}\label{Fes}
F_{\rm es}=\frac{\left(\eps_{_{\rm A}}-\eps_{_{\rm
B}}\right)^2}{2\bar{\eps}}\sum_q (\hat{{\bf
q}}\cdot{\bf E})^2\phi_{\bf q}^2
\end{eqnarray}
where the sum is taken over all $q$-modes in the expression $\phi({\bf
r})=\sum_{\bf q}\phi_{\bf q}\cos({\bf q}\cdot{\bf r})$, and
$\bar{\eps}=f\eps_{_{\rm A}}+(1-f)\eps_{_{\rm B}}$ is the average dielectric
constant. A proper ansatz for the copolymer morphology is a linear combination
of parallel and perpendicular layering: $\phi({\bf r})=w(E)\phi_\parallel({\bf
r})+g(E)\phi_\perp({\bf r})$, with field-dependent amplitudes $w(E)$ and $g(E)$.
When this ansatz is substituted into the free-energy, the amplitudes can be
calculated and the order-parameter obtained.
%
%fig 11
\begin{figure}[h!]
\begin{center}
\includegraphics[scale=0.45,bb=15 435 580 690,clip]{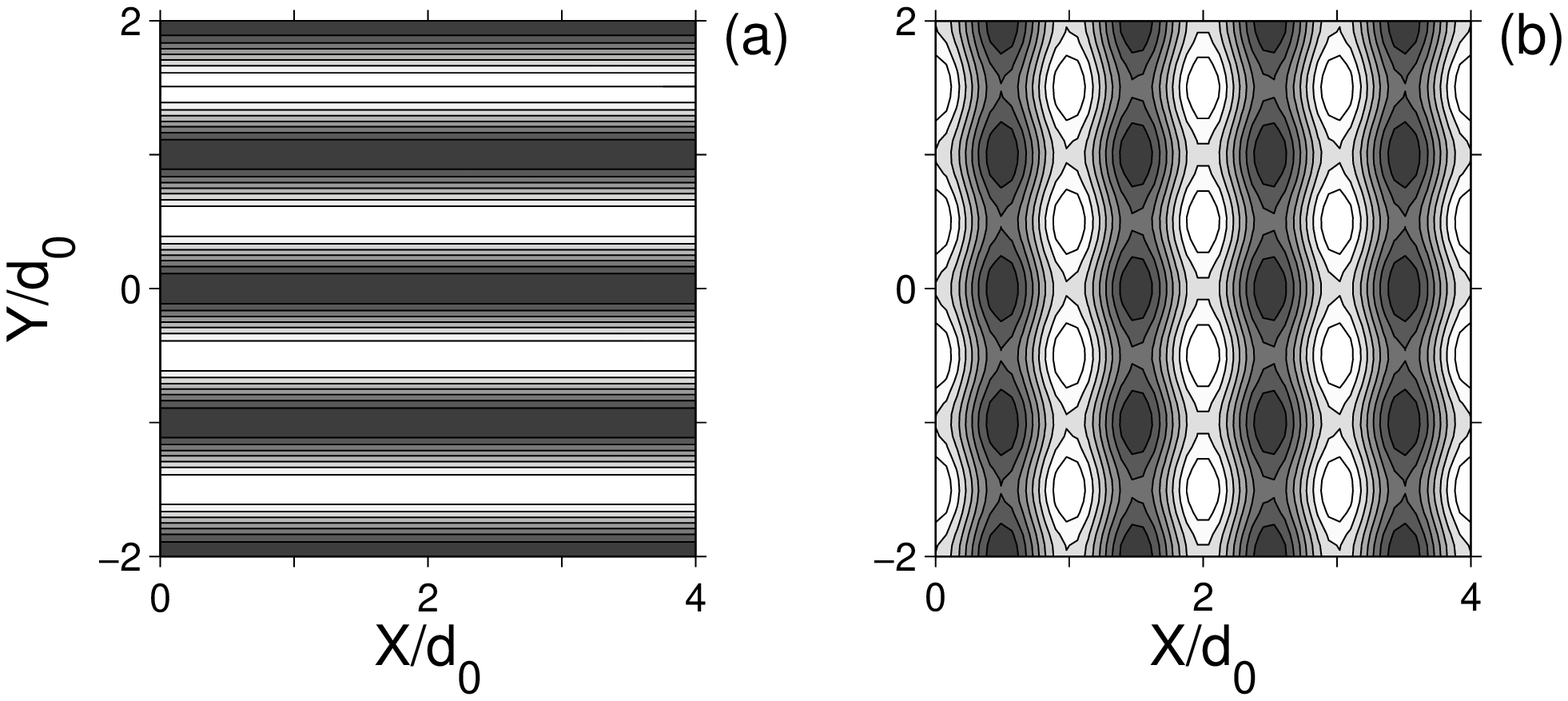}
\end{center}
\caption{\textsf{Lamellar layers of BCP's between two surfaces and
under external electric field. The surfaces are at $y=\pm 2d_0$, and
the field is in the $y$ direction. The B monomers (colored black)
are attracted to both surfaces. (a) The field is slightly smaller
than the critical field, $E=0.98E_c$, and the film has a perfect
parallel ordering. (b) The field is just above the threshold,
$E=1.02E_c$. The film morphology is a superposition of parallel and
perpendicular lamellae. Adapted from ref~\cite{TAmm02}.
%The surface fields are $\sigma^+=\sigma^- =0.5hq_0^3\phi_L$, and the
%Flory %parameter is $N\chi=11$ corresponding to correlation length
%$\xi\simeq 4.8d_0>L$. Adapted from ref~\cite{TAmm02}.
}}
\end{figure}

Figure 11 shows the resulting BCP morphology under external electric
field oriented perpendicular to the surfaces~\cite{TAmm02}. In (a)
the field is just below a {\it critical field} and the lamellae lie
parallel to the electrodes. However, when the field is increased
just above the critical field, a transition occurs to a highly
distorted but predominantly perpendicular layering [see (b)]. As the
field further increases, the modulations diminish and the lamellae
achieve perfect ordering perpendicular to the surfaces.

%fig 12
\begin{figure}[h!]
\begin{center}
\includegraphics[scale=0.7,bb=130 190 450 780,clip]{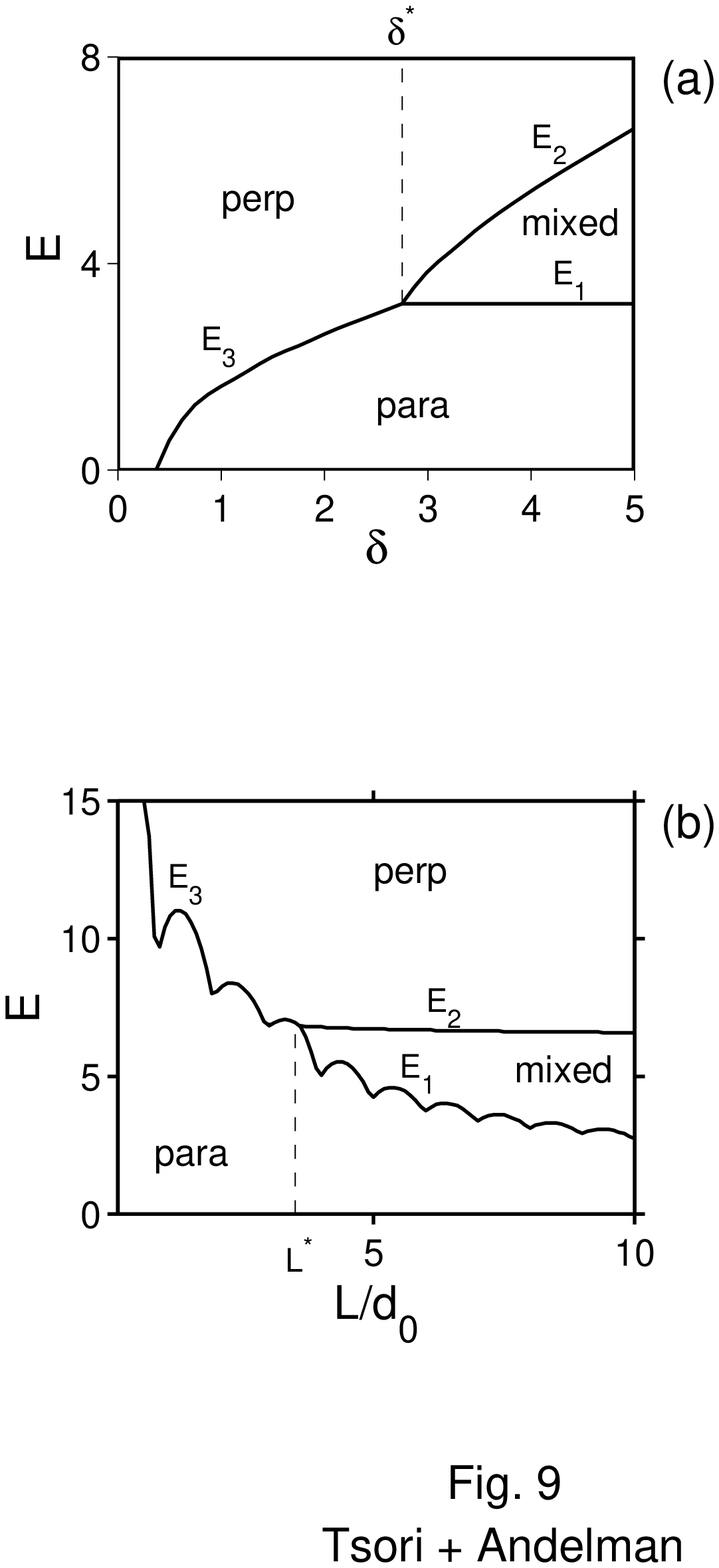}
\end{center}
\caption{\textsf{(a) Phase diagram in the $E$-$\delta$ plane. When
$\delta=(\sigma_{\rm AS}-\sigma_{\rm BS})/\gamma_{_T}<\delta^*$,
there is a transition between parallel and perpendicular lamellae at
$E=E_3$. At larger $\delta$, $\delta>\delta^*$, the transition to
the mixed state is followed by a second transition to the
perpendicular state when $E=E_2$. The surface separation is chosen
as $L=10d_0$. (b) Similar diagram, but in the $E$-$L$ plane, with
$\delta=5$. Adapted from ref~\cite{TAmm02}.}}
\end{figure}

In Fig.~11(b), it is clear that the surface effect propagates far
into the bulk, as the copolymer modulations persist throughout the
whole film. In the strong-segregation regime, however, this is not
true, and the surface effect is localized. We thus can imagine a
third morphology, that of a ``mixed'' phase. In this morphology few
parallel lamellae exist near the surfaces, while the rest of the
film is in the perpendicular orientation. A ``T-junction'' defect is
therefore created, and a surface-tension term $\gamma_{_T}$ must be
associated with it. Indeed such a morphology has been visualized
lately by Russell and co-workers \cite{russell_Tjunction}.

The phase diagram in the plane of $\delta$ and $E$ is shown in
Figure 12 (a), where $\delta\equiv(\gamma_{_{\rm AS}}-\gamma_{_{\rm
BS}})/\gamma_{_T}$, for fixed surface separation~\cite{TAmm02}.
There exist three fields $E_1$, $E_2$ and $E_3$ separating the
parallel, perpendicular and mixed orientations. At small fields,
there is a direct transition from parallel to perpendicular layers
as the field is increased. The mixed state is only possible above a
certain threshold of $\delta$, $\delta^*$. Above this threshold, the
mixed state is stable at fields larger that $E_1$ but smaller than
$E_2$. An increase of $E$ above $E_2$ leads to the stability of
perpendicular lamellae.

The phase diagram in the $L$ and $E$ plane is shown in Figure 12
(b), for a fixed value of $\delta$. At small surface separations, an
increase of $E$ leads to a transition from parallel to perpendicular
lamellae at $E=E_3$. At surface separations larger than a threshold
value $L^*$, increase of $E$ above $E_1$ leads to a mixed
morphology, whereas further increase above $E_2$ gives rise to
perpendicular lamellae.

%%%%%%%%%%%%%%%%%%%%%%%%%%%%%%%%%%%%%%%%%
\subsection{Phase transitions in electric fields}
%%%%%%%%%%%%%%%%%%%%%%%%%%%%%%%%%%%%%%%%%

Orientation of anisotropic phases occurs when the ordered phase has
some freedom to  rotate, and when the applied electric fields are
not too high. However electric fields can cause a phase transition
in systems composed of several components with different dielectric
constants \cite{TTL04}. If the BCP phase under consideration cannot
rotate in order to reduce the electrostatic energy, it begins to
deform. A gradual change then occurs - this is usually an elongation
of domain in the direction parallel to the field. At a certain
point, it is more favorable for the system to make a drastic change
in symmetry and ``jump'' to the state with the best (i.e. minimal)
electrostatic energy \cite{TTL04,prl_ions}. This kind of phase
transition is expected to occur at relatively high fields for BCP's,
$E\sim 50-100$ V/$\mu$m, so sometimes the phase transition is
preempted by dielectric breakdown of the material.

As an example of such a phase transition consider diblock copolymers
in the bcc phase of spheres, under an electric field ${\bf E}$. To
the lowest (quadratic) order in $\phi$, the electrostatic energy is
given by the Amundson-Helfand expression Eq. (\ref{Fes}). Higher
order expressions are available as well \cite{TAES06}. At very high
fields, the spheres will elongate into cylinder oriented along the
field, which can be assumed to lie in the $(1,1,1)$ direction of the
lattice. The transition from perfect bcc to hexagonal symmetries of
$\phi$ can be achieved by the following ansatz \cite{prl_ions}:
\begin{eqnarray}
\phi({\bf r},{\bf
E})=w(E)\sum_{n=1}^{3}\cos({\bf q}_n\cdot{\bf
r})+g(E)\sum_{m=4}^{6}\cos({\bf q}_m\cdot{\bf r})\nonumber\\
\end{eqnarray}
where
\begin{eqnarray}
{\bf q}_{1,4}=q_0 (\mp 1,0,1)/\sqrt{2}\nonumber\\
{\bf q}_{2,5}=q_0 (1,\mp 1,0)/\sqrt{2}\nonumber\\
{\bf q}_{3,6}=q_0 (0,1,\mp 1)/\sqrt{2}
\end{eqnarray}
At zero electric field, $w(E)=g(E)$ and $\phi$ represents a bcc phase.
For large
enough field the order parameter reduces to the hexagonal one:
$g(E)=0$, and $\phi_{\rm hex}({\bf r})=A_{\rm hex}\sum_{n=1}^{3}\cos({\bf
q}_n\cdot{\bf r})$.

%fig 13
\begin{figure}[h!]
\begin{center}
\includegraphics[scale=0.6,bb=50 40 335 745,clip]{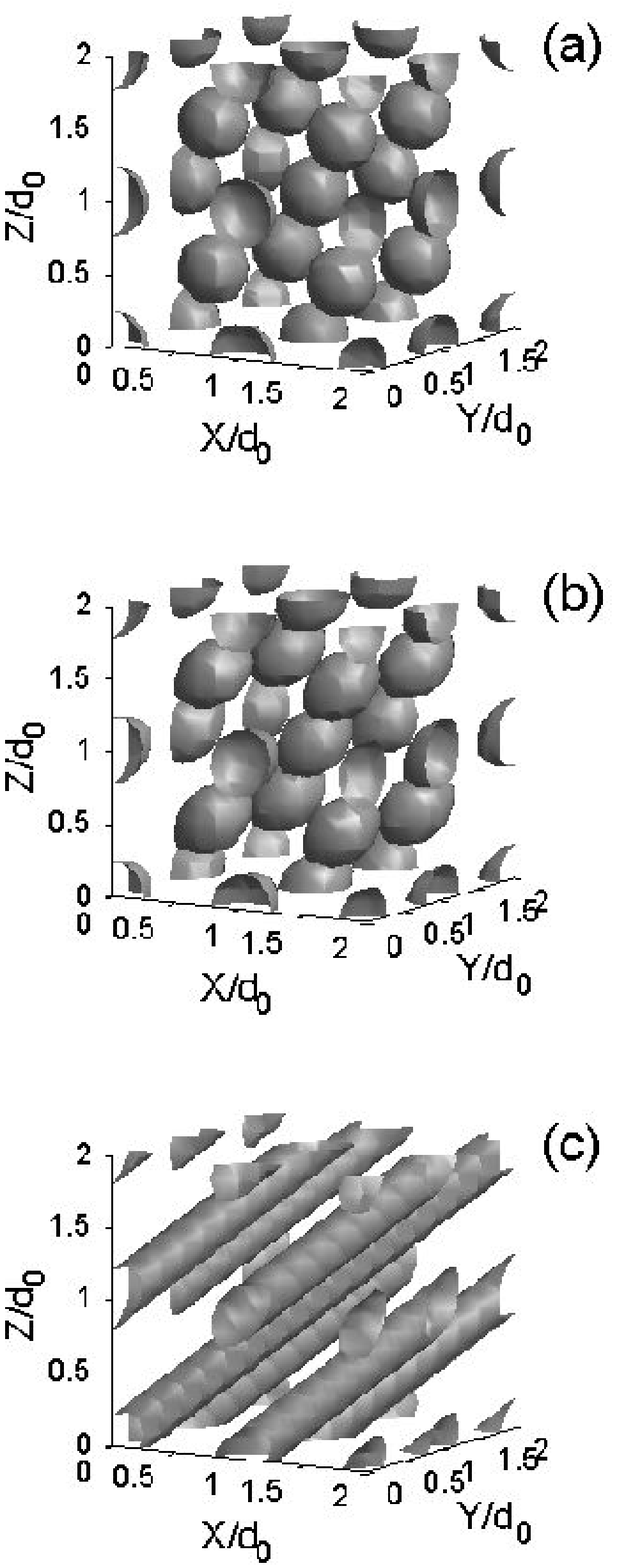}
\end{center}
\caption{\textsf{Cubic phase of block copolymers in electric field.
(a) Electric field is $E=0$. (b) $E=0.98E_c$, just $2$\% below the
critical field, and oriented along the $(1,1,1)$ direction of the
lattice. The spheres are deformed. (c) $E=1.02E_c$, just above the
critical field, and the system undergoes an abrupt change into the
hexagonal array of cylinders. Adapted from ref~\cite{prl_ions}}}
\end{figure}

The result of the minimization of $F=F_b+F_{\rm es}$ gives the value of
$w(E)$ and $g(E)$ and therefore $\phi$. Figure 13 (a) shows the BCP
morphology in the absence of field. As the field is increased, the
spheres deform and elongate along the external field. This state
represents a compromise between electrostatic energy and stretching of
the polymer chains. There exist a critical field $E_c$ above which a
direct transition to cylinders occurs: in (b) the field is $E=0.98E_c$,
while in (c) the field is just above the critical one ($E=1.02E_c$), and
perfect cylinders are formed.

%fig 14
\begin{figure}[h!]
\begin{center}
\includegraphics[scale=0.45,bb=15 180 545 590,clip]{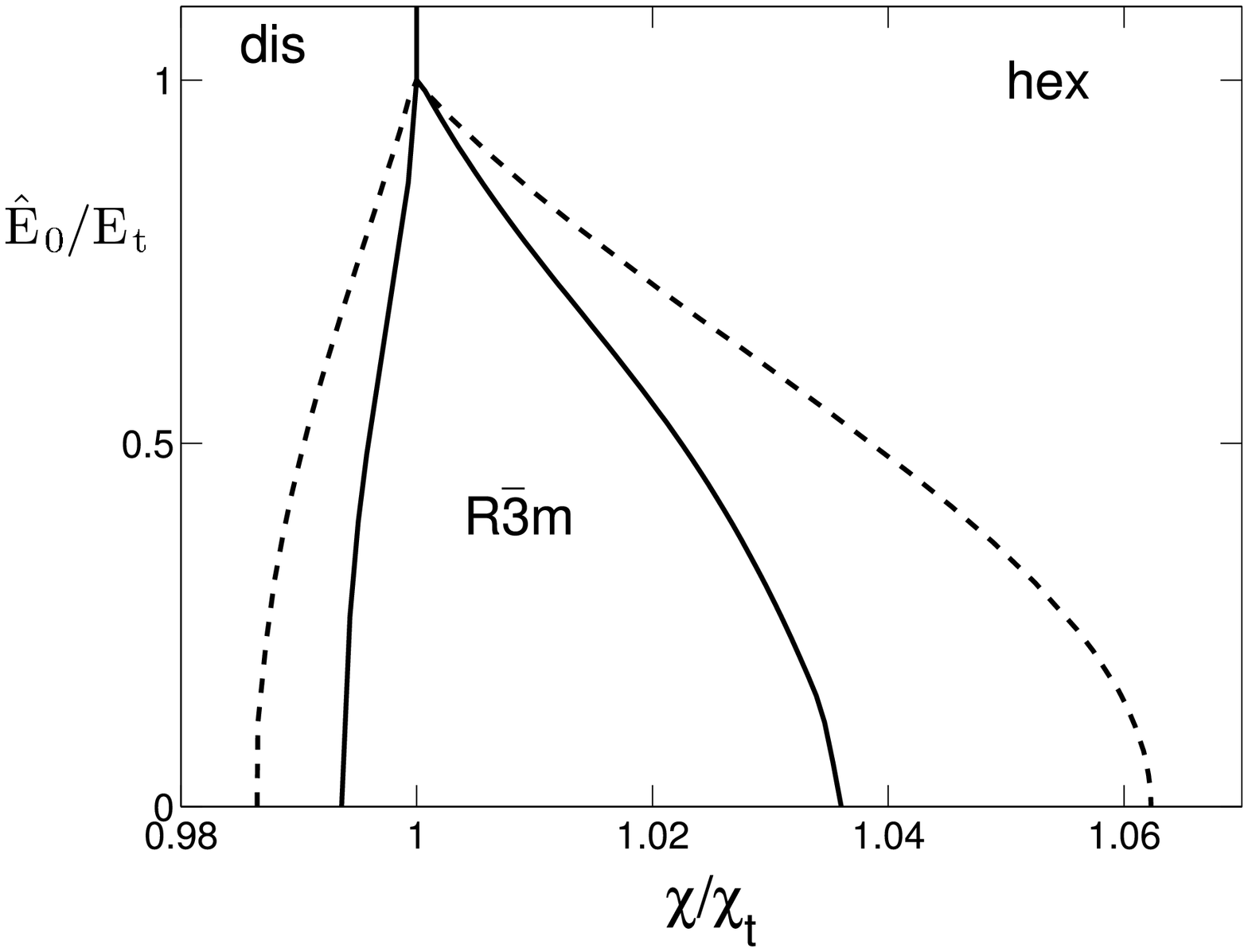}
\end{center}
\caption{\textsf{Phase-diagram of block copolymers in electric
field, in the plane of the Flory parameter $\chi$ and normalized
electric field $\hat{E}_0$. The distorted bcc phase, denoted as
R$\bar{3}$m, is bounded by the hexagonal (hex) and disordered (dis)
phases. Solid line is the prediction of analytical one-mode
approximation, whereas dashed lines are obtained by a more accurate
self-consistent numerical study. Axes are scaled by $(\chi_t,E_t)$,
the values of $\chi$ and $\hat{E}_0$ at the triple point. Adapted
from ref~\cite{TAES06}.}}
\end{figure}

The complete phase diagram taking into account the relative
stability of the various meso-phases can be calculated. We have done
so by two mean-field methods: the first is an analytical method
based on the above coarse-graining approach. The second one is a
more rigorous treatment  based on  numerical solutions of
self-consistent field equations for the copolymer concentration
\cite{TAES06}. The result is shown in Figure 14 in the plane of
field $E$ and Flory parameter $\chi$, and for a particular
composition $f=0.3$. The distorted bcc phases, denoted as
R$\bar{3}$m, is bounded by the hexagonal (hex) and disordered (dis)
phases. All three meet at a triple point
$(E_t,\chi_t)=(0.49,14.11/N)$, where fields are scaled by
$\hat{E}_0\equiv\left(\eps_0v_p/k_BT\right)^{1/2}$, where $\eps_0$
is the vacuum permittivity and $v_p$ is the volume of one copolymer
chain.

%%%%%%%%%%%%%%%%%%%%%%%%%%%%%%%%%%%%%%%%
\subsection{Ionic impurities in BCP's}
%%%%%%%%%%%%%%%%%%%%%%%%%%%%%%%%%%%%%%%%

The above discussion pertains to somewhat ``ideal'' polymers, because in
the electric response only their dielectric constant was considered.
However, most polymers are prepared by anionic polymerization. The
process is initiated by one ButylLithium ion (BuLi). After rinsing with
water, the loose Li bounds with an OH group to form LiOH, some of which
are dissociated. Hence, there is a finite number of positive and negative
ions in the material, and their presence changes the system behavior.

The existence of dissociated ions means that there are additional
{\it forces} which act in the alignment process of the BCP
meso-phases. These forces depend on the mobility $\mu$ of the ions,
and on the frequency $\omega$ of applied field. The torque due to
mobile ions is expected to be large if the drift velocity $\pi e\mu
E/\omega$ is larger than the BCP domain size $~d_0$. In addition,
mobile charges also mean that there is {\it dissipation}. Hence, the
energy stored in the dielectric medium $\eps E^2$ should be compared
to the Joule heating in one cycle of the field $2\pi\sigma
E^2/\omega$, where $\sigma$ is the ions conductivity, proportional
to the ion density. It is clear from the above that for low
frequencies (in practice $\lesssim 100$ Hz) mobile dissociated ions
begin to play an important role in BCP alignment and phase
transitions \cite{mm_ions}.

When the additional complexity due to this ions is taken into account, it
turns out that the orientation forces (torques) due to the mobile ions
scale as $1/\omega$ \cite{mm_ions}, and are equal to the ones due to the
regular ``dielectric mechanism'' at about $50$ Hz. Consequently they are
twice as large at $25$ Hz, and they become more important as the
frequency is reduced.

Taking into account the effect of mobile ions on the BCP phase
transition as outlined above (from bcc phase of spheres to a
hexagonal array of cylinders), it turns out that the transition
field $E_c$ can be significantly reduced from a value of $\approx
70$-$100$\,V/$\mu$m to values $\sim 20$\,V/$\mu$m \cite{prl_ions}.

%%%%%%%%%%%%%%%%%%%%%%%%%%%%%%%%%%%%%%
\section{Conclusions}\label{conclusions}
%%%%%%%%%%%%%%%%%%%%%%%%%%%%%%%%%%%%%%

We review in this paper several ordering mechanisms in confined
BCP's. The theoretical approach relies on a mean-field
coarse-grained Hamiltonian, which is less sensitive to microscopic
details and valid for a wide class of system showing self-assembly
in soft-matter. This approaches thus complements other
computationally-intensive self-consistent numerical schemes
\cite{matsen1,matsen2,pbmm97,shull,MS94} and Monte-Carlo simulations
\cite{binder2,binder3,szleifer1,szleifer2}. The polymer density near
a chemically patterned surface is given above the order-disorder
temperature as a function of the surface pattern. In this regime the
chemical pattern $q$-modes give rise to density modes which are
decoupled from each other (linear response theory).  In the weak
segregation regime, the surface correlations are long range and,
therefore, simple chemical patterns yield complex copolymer
morphology, even though the bulk  is in its disordered phase.

Below  the ODT temperature, we consider lamellae confined by homogeneous
surfaces, and examine the relative stability of parallel vs.
perpendicular ordering as a function of temperature, surface separation
and interfacial interactions. Lamellae confined by striped surface whose
periodicity is larger than the lamellar periodicity appear tilted with
respect to the surface thereby optimizing their surface interactions.
The lamellar undulations are more prominent as the ODT is approached.
Mixed lamellar phases appear when one surface has chemically patterns in
the form of stripes while  the other is uniform.

A  different paradigm for control of BCP orientation in thin films
is using rough surfaces. This method may be advantageous over other
methods in several situations since it is relatively simple to
implement experimentally. The phase diagram is presented for the
ordered phases as a function of surface period, surface amplitude,
and lamellar period, as well as other parameters.

Lastly, the influence of electric field on the phase behavior of
BCP's is considered. An external electric field favors a state where
dielectric interfaces are perpendicular to the field itself. Hence,
lamellae confined in a thin-film are electrostatically preferred
perpendicular rather than parallel to the confining electrodes. In
the weak-segregation regime, there is one critical field at which
parallel layers are transformed into perpendicular ones. Even for
fields larger than the critical field, the long-range effect of the
surfaces is evident as strong lamellar undulations. The strong
segregation regime is considered as well. Here we find three
possible states: parallel, perpendicular and mixed. The last
morphology exhibits few parallel layers close to the electrodes
while the rest of the film is perpendicular. There are either two or
one critical fields separating them, depending on the interfacial
interactions.

An external electric field can also bring about a phase transition
in ordered phases, if they are frustrated and cannot eliminate
dielectric interfaces perpendicular to the field direction. The
transition from the bcc phase of spheres to a hexagonal array of
cylinders under the influence of electric field is discussed as an
example. Below the critical field the spheres elongate in the
field's direction, but above it we find perfect cylinder whose axes
are parallel to the field. The phase diagram of the various
meso-phases is calculated, and the simple analytical expression
obtained with the coarse-graining theory is compared with a more
rigorous SCF theory, with rather good match. We point out that
residual dissociated ions in BCP's can greatly enhance the electric
field effect, and this is specially true in low-frequency electric
fields.

The analytical calculations presented here rely on a relatively
simple mean-field coarse-grained free-energy functional. This
approach allowed us to deal with confinement effects in BCP, take
into account the chemical nature of the surfaces, calculate the
elastic energy penalty and lamellar conformation near curved
interfaces and balance the electrostatic energy against the elastic
one for BCP's in electric fields. The coarse-grained approach has a
big advantage that it can be generalized and account for other
complex polymer systems.  Because it is less accurate in terms of
quantitative predictions, it is useful to compare this approach with
numerical self-consistent field theories, discrete lattice models,
Monte-Carlo and molecular dynamics simulations and experiments.

\bigskip
%%%%%%%%%%%%%%%%%%%%%%%%%%%%%%%%%%%%%%%%
{\it Acknowledgments}
%%%%%%%%%%%%%%%%%%%%%%%%%%%%%%%%%%%%%%%%

~~~ Our research on block copolymer have been conducted in
collaboration with T. Hashimoto, L. Leibler, C.-Y. Lin, M. Schick,
E. Sivaniah and  F. Tournilhac. We would like to thank K. Binder, Y.
Cohen, G. Fredrickson,  S. Gido, G. Krausch, M. Muthukumar, T. Ohta,
G. Reiter, T. P. Russell, U. Steiner, I. Szleifer, E. Thomas, T.
Thurn-Albrecht, M. Turner and T. Xu  for numerous discussions. This
research is partly supported by the Israel Science Foundation (ISF)
under grant No.\ 160/05 and the US-Israel Binational Foundation
(BSF) under grant No.\ 287/02.

\newpage
%------------------------------------------------

\end{document}